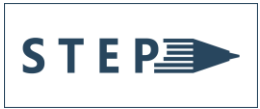
International Journal of Computing Sciences Research (ISSN print: 2546-0552; ISSN online: 2546-115X)
Advance online publication
doi: 10.25147/ijcsr.2017.001.1.115
https://stepacademic.net
Short Paper

# Web-based Database Courses E-Learning Application

Aaron Paul M. Dela Rosa
College of Information and Communications Technology, Bulacan State University, Philippines
aaronpaul.delarosa@bulsu.edu.ph
(corresponding author)

Luigi Miguel M. Villanueva
College of Information and Communications Technology, Bulacan State University, Philippines

John Mardy R. San Miguel
Office of the University Registrar, Bulacan State University, Philippines

John Emmanuel B. Quinto
Technical Department, Infocentric, Philippines
*Date received*: June 22, 2022
*Date received in revised form*: August 12, 2022
*Date accepted*: August 13, 2022

Recommended citation:

Dela Rosa, A. P. M., Villanueva, L. M. M., San Miguel, J. M. R., & Quinto, J. E. B. (2023). Web-based database courses e-learning application. *International Journal of Computing Sciences Research.* Advance online publication. doi: 10.25147/ijcsr.2017.001.1.115
**Abstract**

*Purpose* – This study was focused on the development of a web e-learning application for the database courses taken by Information Technology (IT) students at the College of Information and Communications Technology (CICT) of Bulacan State University (BulSU).

*Method* – The research methodology used in this project was the cross-sectional developmental approach. The Agile Software Development methodology was followed phase by phase, up to the development phase, to develop the system. It was used to produce the desired output rapidly while allowing users to go back through phases without finishing the whole cycle.
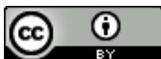
This is an Open Access article distributed under the terms of the Creative Commons Attribution License (http://creativecommons.org/licenses/by/4.0), which permits unrestricted use, distribution, and reproduction in any medium, provided the original work is properly credited.

*Results* – The goal of this study was to create a web application that teaches Structured Query Language (SQL) in both the MySQL and SQL Server approaches. The application contains quizzes and examinations to allow self-assessment of learning. Additionally, an Entity Relationship Diagram (ERD) simulation was included to provide ERD creations in a drag-and-drop fashion. This study was evaluated using ISO/IEC 25010 software quality evaluation criteria. The study's overall mean was 4.24, 4.41, and 4.33, all with the descriptive meaning of *"Very Good,"* which showed that the system performed its necessary functions as perceived by students, faculty members, and experts, respectively.
*Conclusion* – In summary, the e-learning web application for database courses was fully developed. Moreover, the entity-relationship diagram was integrated well within the system and is accessible to the users. Lastly, respondents evaluated the developed web application using the ISO/IEC 25010 with an overall descriptive interpretation of *"Very Good."*

*Recommendations* – For future developments of the study, an administrator panel may be developed to manage users and do other administrative tasks. Additionally, a feature where faculty members may be included to manage contents within the application may be included. Lastly, higher-order thinking skills questions on assessments and quizzes may be included.

*Research Implications* – The e-learning web application was fully developed and may be used as an additional teaching and learning tool upon its implementation as an e-learning web application. Faculty members may use the application as a supplemental tool in teaching the database courses. On the other hand, students may use the web application as a tool in developing entity-relationship diagrams needed for system development courses.

## INTRODUCTION

People can learn, communicate, and play with other people all around the globe. Learning online through technology-based learning can spread knowledge to other people (Stephenson, 2018). Students could be in a class using the internet and can get instant feedback through online learning. Because of that, education online has become popular and beneficial to institutions (Simpson, 2018; Ko & Rossen, 2017). E-Learning is a catch-all term that describes training, or any educational activity delivered to learners through an electronic device (King & Boyatt, 2015). Typically, through a website or an application, the learner accesses a computer (Urh, Vukovic, Jereb, & Pintar, 2015; Islam, Beer, & Slack, 2015). E-Learning web applications are different when students are reading a book. This application can help other people who do not read books anymore and use technology (Arkorful & Abaidoo, 2015; Sauro, 2015).

W3schools is an example of e-learning. This web application allows its users to study and learn anything about developing websites using web programming languages such as JavaScript (JS), Hypertext Preprocessor (PHP), Cascading Stylesheet (CSS), and more (W3Schools, 2022). It provides a part of their learning course where the user may code independently, try out the examples given, and enhance them as to what the user wants. Another excellent web application and example of e-learning for programmers to learn different languages is W3resource (W3resource, 2022). W3resource provides topics about a programming language and provides a complex set of exercises depending on a given topic. Each of the activities has a solution, and the user may code in their solution to a given problem to try it out on their own. Universities may maximize such websites as an additional tool in teaching and learning.

Bulacan State University (BulSU) is a state university located in Malolos, Bulacan, which has 13 colleges on the main campus. The researchers focused on one of the colleges on the main campus, the College of Information and Communications Technology (CICT), which has approximately 2,300 Information Technology (IT) students from 1st to 4th-year levels. 1,000-1,100 students from the total number of IT students at CICT are on their 2nd-year level who are taking database courses. The ratio of the number of students who need to learn Database Management Systems (DBMS) and the number of faculty members who are experts in DBMS may fail. Additionally, with what the pandemic brought to the education sector, maximizing the use of online tools has been beneficial under the flexible learning modality.

The e-learning web application for database courses offered by CICT was developed to address the aforementioned issue. The web application provided lessons about the syllabus's flow in the said college on teaching the database courses. Moreover, this application contains Structured Query Language (SQL) lessons in MySQL and SQL Server. A quiz is given per lesson, and an examination is given per topic. Lastly, the application focuses on discussions, quizzes, and assessments. Faculty members may also use the web application to teach their classes as the web application is delivered following the CICT's course syllabus for these database courses. This application could be used as an additional teaching tool, be it during online or face-to-face classes.

To maximize the development of the e-learning web application, an entity-relationship diagram simulator was included within the application. This feature presents the uniqueness of the developed web application. Using this feature, students could create and develop the said diagram and even relation schemas to be their reference when building a database for system development.

*Project Objectives*

The study aimed to develop an e-learning web application covering database topics provided by Bulacan State University's College of Information and Communications Technology.

Specifically, the following objectives were also considered: (1) to design and develop the web e-learning application of database courses; (2) to integrate an entity-relationship diagram simulation; and (3) to determine the system acceptability using the ISO/IEC 25010 standard software quality evaluation criteria as perceived by the students, faculty members, and IT experts.

**LITERATURE REVIEW**

Encarnacion, Galang, and Hallar (2021) concluded in their study that "e-learning can be considered as one of the best strategies... for teaching and learning." E-learning provides accessibility to learners, enabling them to study even outside of the four corners of the classroom. Moreover, e-learning could provide in-depth discussions and assessments for its learners. Education institutions could strategize its use and integration into the offered courses.

In terms of the stakeholders that will use an e-learning application, Lucero, Victoriano, Carpio, and Fernando (2021) studied the readiness of both students and faculty members. Lucero et al. (2021) concluded that students and faculty members are prepared and ready should e-learning be implemented in educational institutions.

As Encarnacion et al. (2021) identified e-learning as one of the best practices, the development of an e-learning application could be in place and be included as part of the teaching force's additional tools in classes. Additionally, as Lucero et al. (2021) concluded, once the stakeholders are ready, implementing such developed e-learning web applications will be beneficial for both the students and faculty members.

Due to the outbreak of Coronavirus Disease 2019 (COVID-19) beginning in March 2020, education sectors rely on the use of e-learning systems and applications to continue the delivery of education. Zainul et al. (2020) developed Moodle-based e-learning courses with a high level of acceptance from their intended users. Moreover, Irfan, Kusumaningrum, Yulia, and Widodo (2020) maximized the use of e-learning in learning mathematics during the outbreak of the pandemic. Both studies show that even within a pandemic, using an e-learning system could be utilized to continue delivering education to the learners.

The database courses e-learning web application focuses on its goal as an additional tool in teaching and learning. Faculty members may utilize the web application as a supplemental tool in teaching, using its examples in discussions and its available quizzes and assessments. Students have an additional opportunity to learn database courses.

## METHODOLOGY

For the effectiveness and accuracy of the developed system, cross-sectional developmental research was applied. In cross-sectional developmental research, the necessary processes were studied upon developing the web application and evaluating each process to ensure that it shall meet the set criteria for evaluators' acceptance of the developed e-learning web application. Additionally, different age groups and sets of respondents have evaluated the developed system to identify their perception and acceptability of the use of the application (Lumen Learning, 2020; What is Developmental Research?-Definition, Purpose & Methods, 2014). All gathered data was analyzed by its intended users, leading to the development of the database e-learning web application.

### *Project Development*

The system was developed by following the phases of the Agile software development methodology (Kashyap, n.d.; Patel, 2019). It enabled web application development by going through different phases, revisiting a phase if changes to previous phases were required, and modifying contents from previous phases if anomalies were discovered during web application development.

*Requirements* In this phase, the researchers focused on gathering all materials used by CICT in teaching the database courses to be subject to the development of the web application. These identified materials were the syllabi of the two (2) database courses and current instructional materials used by the instructors of these courses.

*Plan.* In this phase, the gathered materials were studied. Additionally, fundamental features of an e-learning application (Sharipovich & O'G'Li, 2020) were taken into consideration. From then, the possible features and their contents were formed and identified for the web application.

*Design.* The focus of this phase is on designing the user interface of the web application. A context diagram was also developed during this phase to identify the user level and which data flows from the entity to the system and vice versa.

*Develop.* After finalizing the requirements of the last phase, the development of the database e-learning web application was started. In this phase, all the features and functionalities of the web application were developed and programmed. Different software development tools were used to achieve the development of the web application. Different web technologies were used to create both the front and back-ends of the e-learning web application. Frameworks, such as Bootstrap and jQuery, were also used as part of the development tools.

*Release, Track, and Monitor*. In the release phase, the system is deployed, operational, and used by its intended users. Following deployment, the tracking and monitoring phase must be completed in order to maintain and update the developed web application. These phases were no longer covered by the study as the study only covers up to the web application development.

## System Evaluation: Software Quality Evaluation Criteria from ISO/IEC 25010:2011

The software quality evaluation criteria from the International Organization for Standardization/International Electrotechnical Commission (ISO/IEC) 25010:2011 (ISO 25000, 2022) were used to determine the acceptability of the respondents to the developed web application. For the sampling techniques used, expert or judgment sampling (Glen, 2015) was used to identify 10 IT experts and professionals who evaluated the web application's acceptability in terms of its technicality. On the other hand, convenience sampling (Maheshwari, 2017) was used on 50 2nd-year IT students of BulSU-CICT. These students were the intended users of the web application. As part of the beneficiaries, convenience sampling (Maheshwari, 2017) was also used on 10 faculty members of BulSU-CICT handling the database courses. A five-point Likert-type scale was used during the evaluation of the aforementioned respondents to the study (Joshi, Kale, Chandel, & Pal, 2015; Willits, Theodori, & Luloff, 2016). Table 1 presents the Likert-type scale used. Upon using the scale, the mean was computed as part of the data presented in the evaluation results.

Table 1. Five-point Likert Scale

| Scale | Range | Descriptive Rating |
|---|---|---|
| 5 | 4.50 – 5.00 | Excellent |
| 4 | 3.50 – 4.49 | Very Good |
| 3 | 2.50 – 3.49 | Good |
| 2 | 1.50 – 2.49 | Fair |
| 1 | 1.00 – 1.49 | Poor |

## RESULTS AND DISCUSSION

### Design and Development of the Web-based Database Course E-Learning Application

The design and development of the web application commenced upon gathering the necessary inputs for the features of the system. A context diagram, also known as a level 0 data flow diagram (DFD), was used to guide the researchers through the process of determining how data can flow as users interact with the web application. Additionally, all features were developed based on what was identified in the context diagram. These features of the web application are the lessons pages (including MySQL and SQL Server discussions), quizzes, and assessment pages.

DFDs show all the data flow from process to process and how to create each function based on the data. The context diagram used as a reference during the web application development is shown in Figure 1.

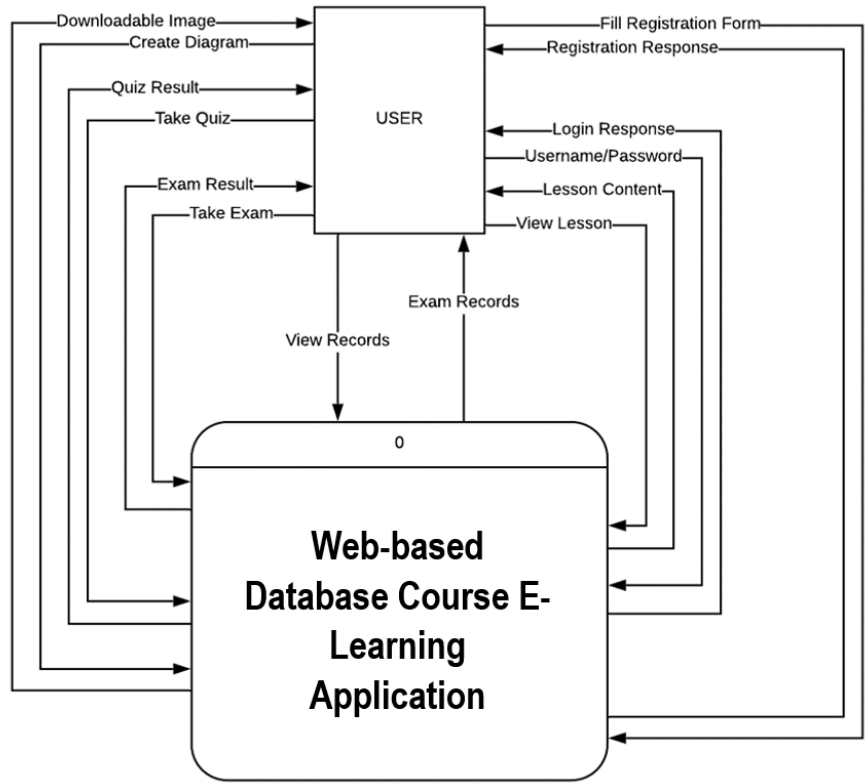

*Figure 1.* Context Diagram (Data Flow Diagram Level 0)

Discussions of MySQL and SQL Server were broken down into chapters that covered everything from the basics to advanced techniques. In terms of the lessons provided within the web application, Figure 2 shows the web application's lesson portion.

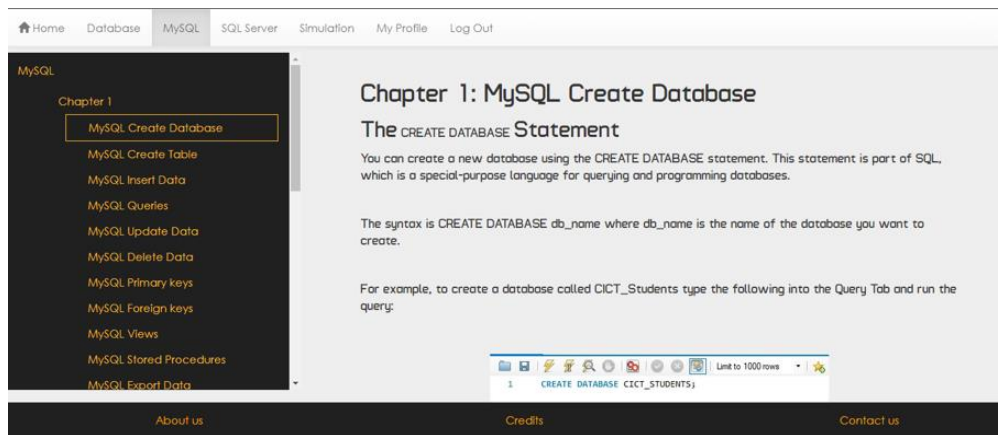
*Figure 2.* Lesson page

On the other hand, Figure 3 shows the web application in terms of the example code provided within the web application.

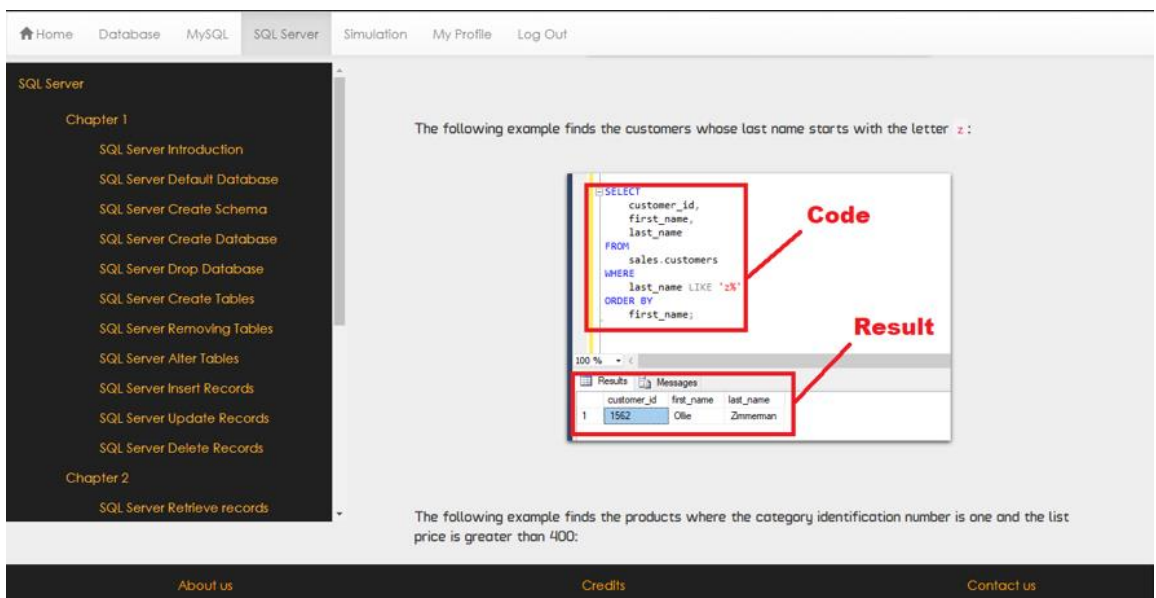
*Figure 3.* SQL Code and Results in a Lesson page

### *Integration of Entity-Relationship Diagram Simulation*

An Entity-Relationship Diagram (ERD) Simulator was integrated, which is used in creating ERDs. Figure 4 shows the simulation page where the users may select which element should appear in the diagram. The components may be moved through a drag-and-drop action to arrange the diagram according to their liking. Moreover, they may create a relational schema using the simulation. Figure 5 shows how users may create a relational schema. Options for the elements were provided at the top of the workspace.

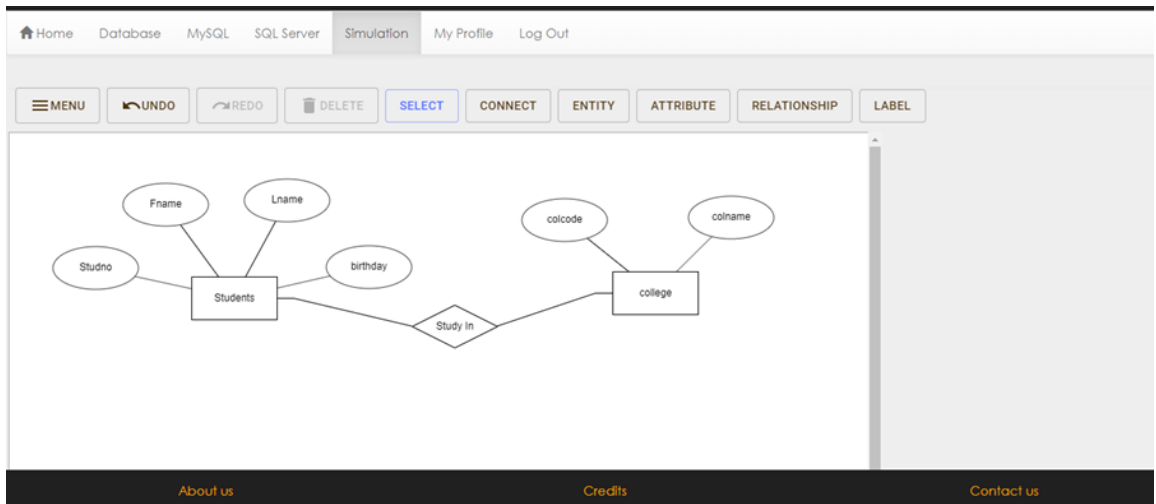
*Figure 4.* Simulation on developing an ERD

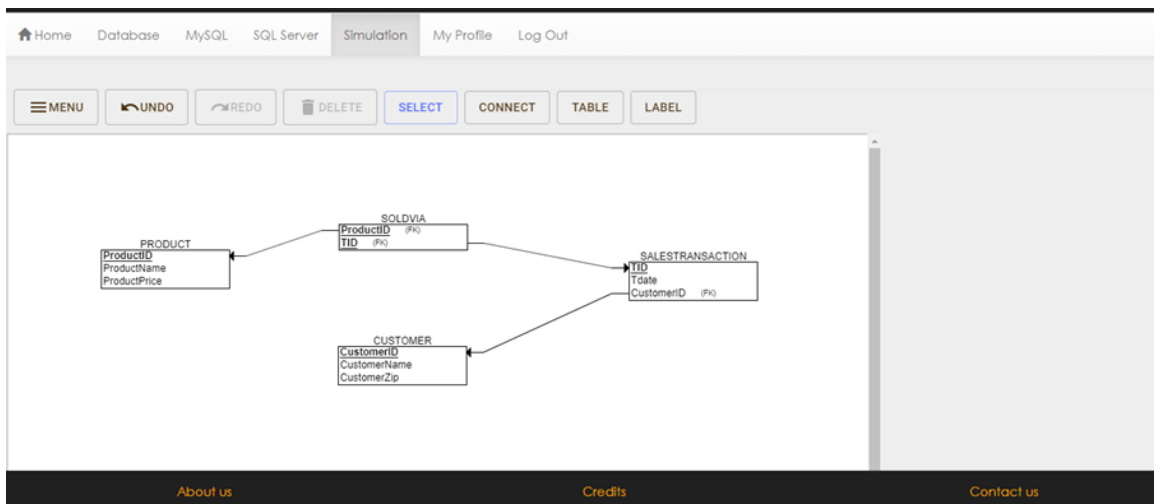
*Figure 5.* Simulation on creating Relational Schema

## System Acceptability Using the Software Quality Evaluation Criteria from ISO/IEC 25010:2011

The developed database e-learning web application was evaluated by students, faculty members, and experts to measure the system's acceptability. The system's acceptability was calculated from the model of ISO/IEC 25010:2011 (ISO 25000, 2022) with the following criteria: Functional Suitability, Performance Efficiency, Compatibility, Usability, Reliability, Security, Maintainability, and Portability. Table 2 presents the overall summary of the respondents' ratings as perceived by the students, faculty members, and technical experts of the web application.

Table 2. Overall Summary of the Respondents' Ratings to the Web Application

| Item | Student Mean | Student DI | Faculty Mean | Faculty DI | IT Experts Mean | IT Experts DI |
|---|---|---|---|---|---|---|
| Functional Suitability | 4.37 | Very Good | 4.47 | Very Good | 4.42 | Very Good |
| Performance Efficiency | 4.33 | Very Good | 4.47 | Very Good | 4.46 | Very Good |
| Compatibility | 4.34 | Very Good | 4.60 | Excellent | 4.36 | Very Good |
| Usability | 4.21 | Very Good | 4.50 | Excellent | 4.25 | Very Good |
| Reliability | 4.16 | Very Good | 4.25 | Very Good | 4.35 | Very Good |
| Security | 4.04 | Very Good | 4.24 | Very Good | 4.14 | Very Good |
| Maintainability | 4.15 | Very Good | 4.28 | Very Good | 4.28 | Very Good |
| Portability | 4.33 | Very Good | 4.47 | Very Good | 4.41 | Very Good |
| **Overall Mean** | **4.24** | **Very Good** | **4.41** | **Very Good** | **4.33** | **Very Good** |

*DI = Descriptive Interpretation*

Among the three groups, the faculty members gave the highest rating, with an overall mean of $\bar{x}$=4.41 and a descriptive interpretation of *Very Good*. This indicates that the faculty members deemed that the application, upon its implementation, could be useful as an additional teaching tool as it comprised two (2) relational database management systems (RDBMSs) in its lessons and discussions. The students rated functional suitability the highest ($\bar{x}$=4.37) with a descriptive interpretation of *Very Good*, which indicates that the features and functionalities provided within the e-learning web application are well developed to its appropriate functionality. In addition, the same criterion was rated the second highest by the experts ($\bar{x}$=4.42). Lastly, the experts rated performance efficiency the highest ($\bar{x}$=4.46) with a descriptive interpretation of *Very Good*, which indicates that the experts found that the web applicate performs efficiently when accessing its contents. Moreover, the web application performed well in its simulation and creation of the ER diagrams and relational schemas.

## CONCLUSIONS AND RECOMMENDATIONS

In summary, the e-learning web application for database courses was fully developed, providing necessary features of an e-learning application such as the provision of lessons, assessments, discussions, and quizzes. Moreover, the entity-relationship diagram was integrated well within the system and is accessible by the users to create and develop ERDs and relational schemas. Lastly, respondents evaluated the developed web application using the ISO/IEC 25010 Students, faculty members, and IT experts rated the developed web application with an overall descriptive interpretation of "*Very Good*," which shows that the application functions well and provides correct data presentation.

The following recommendations were made considering the summary of findings of the study: The goals are to (1) develop an administrator panel that could manage users and do other administrative tasks; (2) develop a feature where faculty members are able to manage lessons, discussions, quizzes, and assessments within the application; and (3) include higher-order thinking skills questions on assessments and quizzes to test the user's knowledge of complexity since this application is more self-paced learning.

## IMPLICATIONS

The e-learning web application was completed and will be used as an additional teaching and learning tool once it is implemented. Faculty members may use the application as a supplemental tool in teaching the database courses to inculcate additional learning and examples for their students. Additionally, assessments and quizzes provided within the application may be used by the faculty members and be included in the course requirements of the database courses taught at BulSU-CICT. On the other hand, students may maximize the web application to create and develop entity-relationship diagrams and relational schemas needed for their system development courses.

## ACKNOWLEDGEMENT

This work does not receive any funding upon its development. The researchers would like to thank Bulacan State University for the presentation of this paper at an international conference. Gratitude shall also be extended to Ralph David Rafael, John Ronel De Leon, Jay Leonard Ruiz, Jan Louie Herrera, and Mark Angelo Zuñiga for their untiring support during the development of this study.

Learning' with Moodle-based for prospective teacher in Indonesia. *Journal of Physics: Conference Series*, 1594, 1-8. doi:10.1088/1742-6596/1594/1/012023

**Authors' Biography**

Aaron Paul M. Dela Rosa, MSIT, is currently a student in the Doctor of Information Technology program and is currently working on his dissertation. Mr. Dela Rosa is a college instructor at the College of Information and Communications Technology (CICT) of Bulacan State University (BulSU). He is also an Online Learning Environment Specialist under the Educational Development Office (EDO) of the Office of the Vice President for Academic Affairs (OVPAA). He presents research papers at national and international conferences focusing on web application development and related fields. He is a member of multiple national and international organizations, and he is a certified data protection officer and certified in multiple programming languages.

Luigi Miguel M. Villanueva is a graduate of the Bachelor of Science in Information Technology at Bulacan State University. Currently, Mr. Villanueva owns businesses where he practices his degree as one of the major designers.

John Mardy R. San Miguel is a graduate of the Bachelor of Science in Information Technology at Bulacan State University. Mr. San Miguel is currently a programmer/web developer in the Office of the University Registrar (OUR) in BulSU. He is also proficient in different web development tools, which he is currently using to create the OUR website.

John Emmanuel B. Quinto is a graduate of the Bachelor of Science in Information Technology at Bulacan State University. Mr. Quinto is now working as a solutions engineer in the technical department of Infocentric. He is also a Certified Cisco Network Associate (CCNA), an Alibaba Cloud Associate (ACA), and an Aruba Certified Switching Associate (ACSA).